\documentclass{article}

\usepackage{amssymb,graphicx}
\usepackage{pstricks} 

\newcommand{\Pythia}{\textsc{Pythia}}
\newcommand{\Jetset}{\textsc{Jetset}}
\newcommand{\Fritiof}{\textsc{Fritiof}}

\begin{document}
\title{High energy electroproduction off complex nuclei}

\author{T.~Falter, W.~Cassing, K.~Gallmeister, and U.~Mosel\\
\small{\it Institut f\"ur Theoretische Physik}\\ 
\small{\it Universit\"at Giessen}\\ 
\small{\it Germany}}

\date{April 20, 2005}

\maketitle

\begin{abstract}
We investigate hadron attenuation in deep-inelastic lepton scattering off complex nuclei in the kinematic regime of the HERMES experiment. Our transport theoretical simulations reveal strong prehadronic final state interactions of the reaction products with the surrounding nuclear medium early after the initial $\gamma^*N$ interaction has taken place. In this work we compare our model results with the measured hadron multiplicity ratios for a $^{84}$Kr target at HERMES and provide an extended discussion of the double-hadron attenuation recently observed at HERMES.
\end{abstract}
\bigskip

\section{Introduction} \label{sec:intro}
Hadron production in high energy lepton scattering off nuclear targets provides an ideal tool to study the physics of hadronization \cite{Kop}. In elementary electron-proton reactions nothing can be learned about the space-time picture of hadron formation, since the reaction products hadronize long before they reach the detector. In nuclear reactions the nucleus serves as a kind of micro-detector which is located directly behind the primary $\gamma^*N$ interaction point. The attenuation of produced hadrons in the nucleus compared to a proton or deuterium target then yields combined information on the length scales of hadron formation and the strength of the prehadronic final state interactions (FSI). 

In the recent past the HERMES collaboration \cite{HERMESDIS,HERMESDIS_new} has carried out a detailed study of hadron production in deep inelastic lepton scattering on various gas targets using the 27.6 GeV positron (electron) beam at DESY. The observed attenuation of produced hadrons has led to two complementary interpretations: In Refs.~\cite{ArleoWang} it is assumed that hadronization occurs far outside the nucleus and that the attenuation of high energy hadrons is solely caused by a partonic energy loss. Here the struck quark undergoes multiple scattering while propagating through the nucleus giving rise to induced gluon radiation. On the other hand Refs.~\cite{Accardi,Falteralt,Falterneu,Kopeliovich} are based on the idea that the struck quark forms a color neutral prehadron early after the initial $\gamma^*N$ interaction and that the prehadronic FSI modify the multiplicity spectra of hadrons. The authors of Refs.~\cite{Accardi,Kopeliovich} treat the FSI purely absorptive and restrict their considerations to direct interactions of the virtual photon with valence quarks inside the nucleus. In Refs.~\cite{Eff00,Falter} we have developed a method to incorporate quantum mechanical coherence length effects observed in high-energy photon-nucleus reactions into our semi-classical BUU transport model. This enables us to account for direct {\it and} resolved photon-nucleus interactions and to provide a probabilistic coupled-channel treatment of the (pre-)hadronic FSI. Furthermore, our Monte Carlo approach allows for realistic event-by-event simulations that can either be corrected for detector efficiencies or which can directly be used as input for detector simulations. In Refs.~\cite{Falteralt,Falterneu} we have applied our model to hadron attenuation at HERMES and EMC energies and in Ref.~\cite{FalterJLab} we also made first predictions for the kinematic regime of the current JLab experiments.

Besides making an essential contribution to our understanding of hadronization, electroproduction off complex nuclei also provides useful input for the interpretation of observables in ultra-relativistic heavy ion collisions. In the latter case the system passes through various energy and density phases and the collision geometry is much less under control than for lepton-nucleus scattering where the target remains more or less in its ground state throughout the reaction. The hadron formation lengths extracted from electron-nucleus reactions help to clarify to what extend the recently observed quenching of high transverse momentum hadrons in ultra-relativistic heavy ion collisions at RHIC is due to partonic energy loss in a quark gluon plasma \cite{Gal03,Cass04}.

\section{Model}
In our model we split the lepton-nucleus reaction into two parts: In the first step the exchanged virtual photon interacts with a bound nucleon inside the nucleus, thereby producing a prehadronic final state which -- in step two -- is propagated in our BUU transport model. 

\begin{figure}[tb]
\begin{center}
\includegraphics[scale=1.]{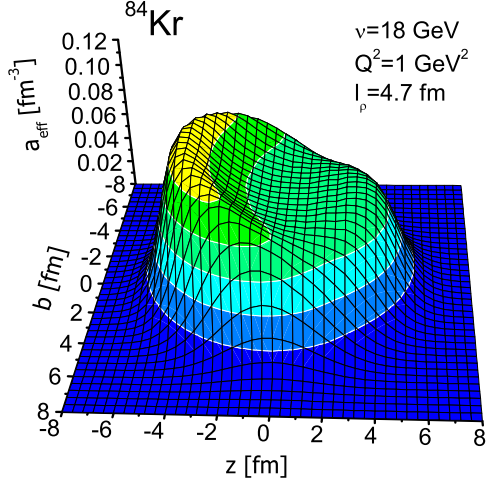}
\end{center}
\vspace{-1.5cm}
\caption[]{Profile function for shadowing. Shown is the probability
distribution for the interaction of an incoming photon (from left)
with given virtuality $Q^2$ and energy $\nu$ with nucleons in a $^{84}$Kr
nucleus.}
\label{fig:profile}
\end{figure}

In the initial $\gamma^*N$ interaction we account for nuclear effects such as binding energies and Fermi motion of the bound nucleons as well as Pauli blocking for outgoing fermions. Furthermore, we calculate the wave function modifications of the resolved photon components, i.e.~the vector meson and perturbative $q\bar{q}$-fluctuations, on the path through the nucleus up to the primary production point. As a result the nucleons on the front side of the nucleus shadow the downstream nucleons from resolved photon interactions at large coherence lengths \cite{Eff00,Falter}. The situation is depicted in Fig.~\ref{fig:profile} for a $^{84}$Kr nucleus and a high-energy virtual photon with momentum directed along the positive $z$-axis and a $\rho^0$ coherence length of the order of the nuclear radius. 

The product of the high-energy $\gamma^*N$ interaction is determined by the event generators \Pythia{} \cite{PYTHIA} and \Fritiof{} \cite{FRITIOF} which allows us i) to account for direct interactions of the bare virtual photon with sea and valence quarks inside the nucleon and ii) also for the resolved photon interactions. In the latter case the hadronic fluctuations of the photon can participate in a hard or a diffractive scattering process with the nucleon. All scatterings lead to the excitations of hadronic strings which subsequently decay according to the Lund fragmentation scheme \cite{Lund} into color neutral prehadrons. In this work we assume that these prehadrons are produced directly after the $\gamma^*N$ interaction. This is a simplifying assumption that is generally used in transport models \cite{Eff00,HSDUrQMD}. The prehadrons then need a formation time to build up their hadronic wave function and to interact like physical hadrons. For simplicity we take the formation time to be a universal constant $\tau_f=0.5$ fm/$c$ in the rest frame of each hadron. This corresponds approximately to the time that the hadron constituents need to travel a distance of a typical hadronic radius. Furthermore, it is in agreement with formation times that have been extracted from particle production in $pA$ collisions at AGS energies within a similar transport model \cite{Cas02}. During their formation time the prehadrons are assumed to interact with reduced cross sections which are determined by the number $n_\mathrm{org}$ of {\it leading} (anti-)quarks that were initially present either in the target nucleon or the hadronic fluctuation of the photon:
\begin{eqnarray}
\label{eq:prehadrons} \sigma^*_\mathrm{prebaryon}&=
&\frac{n_\mathrm{org}}{3}\sigma_\mathrm{baryon} , \nonumber\\
    \sigma^*_\mathrm{premeson}&=&\frac{n_\mathrm{org}}{2}\sigma_\mathrm{meson}\,.
\end{eqnarray}
Consequently, the hadrons that contain leading quarks, i.e. the beam and target remnants, interact with a (reduced) cross section right after the $\gamma^*N$ interaction while the prehadrons that solely contain (anti)quarks produced from the vacuum in the string fragmentation will start to interact after their formation time $t_f=\gamma\tau_f$.

The propagation of the final state (pre)hadrons through the nucleus is described by our BUU transport model. The model is based on a set of generalized transport equations which determines the time evolution of the spectral phase-space distributions $F_h$ of the (pre)hadrons. The distributions $F_h(\vec r, \vec p,\mu, t)$ give at each moment of time $t$ and for each particle class $h$ the probability to find a particle of that class with a (possibly off-shell) mass $\mu$ and momentum $\vec p$ at position $\vec r$. Its time-development is given by the equation
\begin{equation}
	\label{eq:BUU}
	\left(\frac{\partial}{\partial t}+\frac{\partial H_h}{\partial \vec p}\frac{\partial}{\partial \vec r}-\frac{\partial H_h}{\partial \vec r}\frac{\partial}{\partial \vec p}\right)F_h=G_h\mathcal{A}_h-L_hF_h
\end{equation}
where $H_h$ denotes the relativistic energy of the (pre)hadron $h$ and contains a momentum- and density-dependent mean-field potential in case of baryons. The terms on the lhs of ($\ref{eq:BUU}$) are the so-called {\it drift terms} since they describe the independent transport of each hadron class $h$. The terms on the rhs of ($\ref{eq:BUU}$) are the {\it collision terms}; they describe both elastic and inelastic collisions between the hadrons. Here the term inelastic collisions includes those collisions that either lead to particle production or particle absorption. The former is described by the gain term $G_h\mathcal{A}_h$ on the rhs in ($\ref{eq:BUU}$), the latter process (absorption) by the loss term $L_hF_h$. Note that the
gain term is proportional to the spectral function $\mathcal{A}_h$ of the particle being produced, thus allowing for production of off-shell particles. On the contrary, the loss term is proportional to the spectral phase space distribution itself: the more particles there are the more can be absorbed. The terms $G_h$ and $L_h$ on the rhs give the actual strength of the gain and loss terms, respectively. They have the form of Born-approximation collision integrals and take the Pauli-principle into account. The free collision rates themselves are taken from experiment or are calculated \cite{Falterneu,Eff99c}.

\section{Results}
In Ref.~\cite{Falterneu} we have presented a detailed comparison of our model results and experimental data from the HERMES experiment. In this work we review our major findings.

In our simulations we apply the same kinematic cuts as in the HERMES experiment and in addition account for the geometrical acceptance of the HERMES detector. The first observable that we discuss is the so called multiplicity ratio
\begin{equation}
\label{eq:multiplicity-ratio}
R_M^h(z_h,\nu,p_T^2)=\frac{\frac{N_h(z_h,\nu,p_T^2)}
{N_e(\nu)}\big|_A}{\frac{N_h(z_h,\nu,p_T^2)}{N_e(\nu)}\big|_D}
\,,
\end{equation}
where $N_h$ is the yield of semi-inclusive hadrons in a given $(z_h,\nu,p_T^2)$-bin and $N_e$ the yield of inclusive deep inelastic scattering leptons in the same $\nu$-bin. Like in experiment we define the hadron fractional energy as $z_h=E_h/\nu$ and denote the transverse hadron momentum component with respect to the virtual photon momentum as $p_T$. For the deuterium target, i.e.~the denominator of Eq.~(\ref{eq:multiplicity-ratio}), we simply use the isospin averaged results of a proton and a neutron target. Thus in the case of deuterium we neglect the FSI of the produced hadrons and also the effect of shadowing and Fermi motion. The latter will be accounted for in an upcoming work.

\begin{figure}[h!]
\begin{center}
    \includegraphics[scale=1.2]{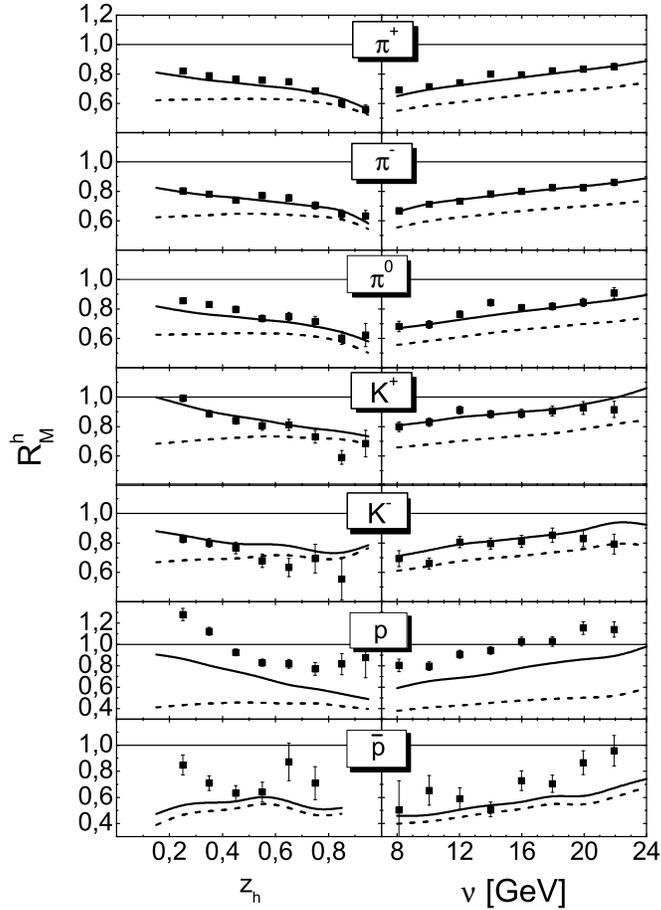}
\end{center}
	\vspace{-0.5cm}
  \caption[]{Multiplicity ratios of $\pi^{\pm,0}$, $K^\pm$, $p$ 
  and $\bar{p}$ for a $^{84}$Kr nucleus (when using a 27.6 GeV
  positron beam at HERMES) as a function 
  of the hadron energy fraction $z_h=E_h/\nu$ and the photon energy 
  $\nu$. The solid line represents the result of a simulation, 
  where we use the constituent quark concept (\ref{eq:prehadrons}) 
  for the prehadronic cross sections and a formation time 
  $\tau_f=0.5$ fm/$c$. The dotted line
  indicates the result of a simulation where the FSI have been treated
  purely absorptive. The data are taken from 
  Ref.~\cite{HERMESDIS_new}.}
  \label{fig:Krid}
\end{figure}
Fig.~\ref{fig:Krid} shows the multiplicity ratio (\ref{eq:multiplicity-ratio}) for the production of pions, (anti-)kaons and (anti-)protons on a $^{84}$Kr target. The solid line represents the result of our coupled-channel BUU calculation using an universal formation propertime $\tau_f=0.5$ fm/$c$ and our simple constituent quark concept (\ref{eq:prehadrons}). As we have discussed in Ref.~\cite{Falterneu} the increase of $R_M^h$ with photon energy $\nu$ is not solely caused by an increase of time dilatation but may in parts be due to the geometrical acceptance of the HERMES detector which is restricted to the forward region. The increase of hadron attenuation with rising fractional hadron energy $z_h$ arises from the fact that a high $z_h$ (pre)hadron most likely contains a leading quark and therefore might get absorbed early after the $\gamma^*N$ interaction. Furthermore, the elastic and inelastic FSI shuffle strength from higher to lower (fractional) energies as can be seen by comparing to a calculation without coupled-channel treatment (dashed line). In the latter case we simply remove all particles that have undergone a collision from the outgoing channel. As expected the discrepancy is largest at small $z_h$. One also sees that the coupled-channel treatment leads to an increasing yield of low energy protons that are knocked out of the nucleus in the FSI. However, the kinematic cuts reduce this increase in such a way that we still overestimate the attenuation of protons.

From Fig.~\ref{fig:Krid} one sees that the absorption strength is about the same for kaons and anti-kaons. This is due to the following effects: Because of reactions like $\bar{K}N\to\Lambda\pi\ldots$ the absorption cross section of $K^-$ is larger than that of $K^+$. This is partly compensated by a reduction of the prehadronic interactions. This is due to a smaller probability to form a leading $K^-(\bar{u}s)$ in the primary $\gamma^*N$ interaction, since now the photon has either to interact directly with a sea quark inside the nucleon or undergo a resolved interaction. Furthermore, there exists the possibility to create $K^+(u\bar{s})$ through reactions like $\pi N\to\Lambda K\ldots$ in the FSI.

\begin{figure}[tb]
\begin{center}
    \includegraphics[width=8cm]{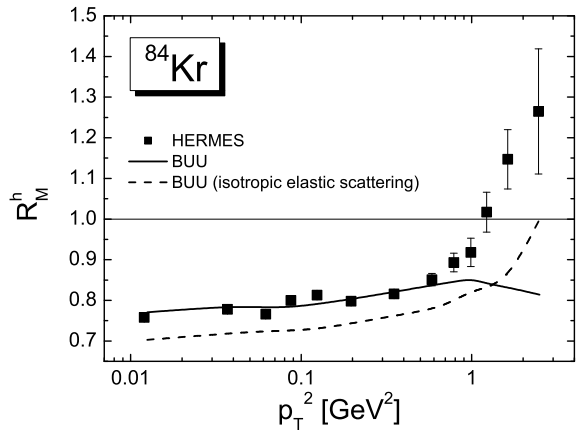}
\end{center}
  \caption{The multiplicity ratio of charged hadrons for $^{84}$Kr (at HERMES) as a function of the transverse momentum squared $p_T^2$. In the simulation we use the constituent quark concept (\ref{eq:prehadrons}) for the prehadronic cross sections and a formation time $\tau_f=0.5$ fm/$c$. In the simulation indicated by the dashed line we additionally assume that all elastic scatterings are isotropic in their center-of-mass frame. The data are taken from Ref.~\cite{HERMESDIS_new}.}
   \label{fig:Krpt}
\end{figure}

The HERMES collaboration has also measured the $p_T$-dependence of the multiplicity ratio (\ref{eq:multiplicity-ratio}). The result is shown in Fig.~\ref{fig:Krpt} for a $^{84}$Kr target in comparison to our BUU calculation indicated by the solid line. Obviously, the experimental multiplicity ratio shows a strong increase at transverse momenta above 1~GeV which can be seen as the Cronin effect in $eA$ collisions and which is not reproduced by our (pre)hadronic FSI. In addition to our standard approach -- which uses the vacuum cross sections as input for determining the outcome of the rescatterings in the FSI -- we have also performed a calculation where we assumed all elastic scatterings in the FSI to be isotropic (dashed line in Fig.~\ref{fig:Krpt}). The latter leads to an increase of the theoretical multiplicity ratio at high transverse momenta, however it overestimates the total strength of the attenuation! The Cronin effect in $eA$ collisions therefore should have a partonic origin which might either be due to partonic rescattering in the FSI, an increase of the intrinsic transverse momentum $\langle k_T\rangle$ of the resolved photon components on their way through the nucleus -- an initial state effect -- or an increase of $\langle k_T\rangle$ in the bound target nucleons.

Finally, we discuss the double-hadron attenuation that has recently been measured at HERMES \cite{Nezza}. In each event only the two (charged {\it or} neutral) hadrons with the highest energies are considered. In the following we denote the hadron with the highest $z_h$ as the leading hadron and the other one as the subleading hadron. The experimental observable is the double-hadron attenuation ratio
\begin{equation}
\label{eq:double-hadron}
    R_2(z_2)=\frac{\frac{N_2(z_2)}{N_1}\big|_A}{\frac{N_2(z_2)}{N_1}\big|_D}.
\end{equation}
Here $N_2(z_2)$ denotes the number of events where the leading and subleading hadron carry the energy fraction $z_1>0.5$ and $z_2<z_1$, respectively, and $N_1$ is the number of events where at least one of them has $z_h>0.5$. Again we account for all experimental cuts and the geometrical acceptance of the HERMES detector in our simulation.

\begin{figure}[tb]
\begin{center}
    \includegraphics[width=6cm]{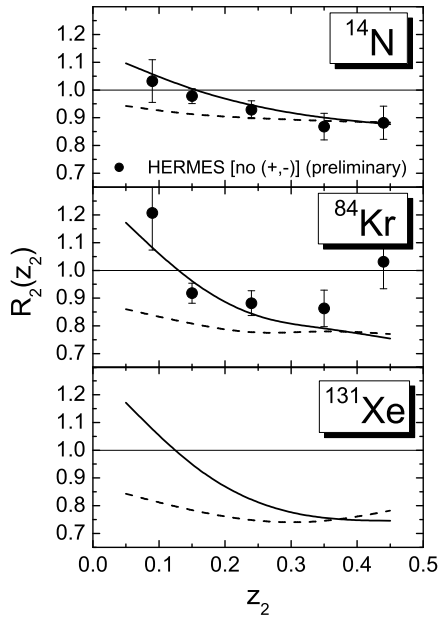}
\end{center}
  \caption{Double-hadron attenuation ratio $R_2$ for a $^{14}$N, $^{84}$Kr and $^{131}$Xe target as a function of the energy fraction $z_2$ of the subleading hadron. In the simulation (solid line) we use the constituent quark concept (\ref{eq:prehadrons}) for the prehadronic cross sections and a formation time $\tau_f=0.5$ fm/$c$. To exclude contributions from $\rho^0$ decay into $\pi^+\pi^-$ the charge combinations '$+-$' and '$-+$' have been excluded both in experiment and in the simulation. The dashed line shows a calculation with a purely absorptive treatment of the FSI. The preliminary HERMES data are taken from Ref.~\cite{Nezza}.}
   \label{fig:double}
\end{figure}

Fig.~\ref{fig:double} shows our model prediction for the double-hadron multiplicity ratio (\ref{eq:double-hadron}) for $^{14}$N, $^{84}$Kr and $^{131}$Xe. To exclude contributions from $\rho^0$ decay into $\pi^+\pi^-$ the charge combinations '$+-$' and '$-+$' have been excluded both in experiment and in the simulation. The solid line shows the result of a full coupled-channel calculation using the constituent quark concept (\ref{eq:prehadrons}) and the formation time $\tau_f=0.5$ fm/$c$. The shape of the spectrum is similar to that of $R_M^h(z_h)$ of the single hadron multiplicity ratio shown in Fig.~\ref{fig:Krid}. As can be seen from Fig.~\ref{fig:double}, our calculations predict about the same double-hadron attenuation ratio for $^{131}$Xe and $^{84}$Kr. As we explain below the attenuation of leading and double hadrons increases in the same way when going from the krypton to the xenon target. Hence, the double-hadron attenuation ratio (\ref{eq:double-hadron}) stays roughly the same. Note, that this does not necessarily imply the same single-hadron attenuation for $^{84}$Kr and $^{131}$Xe. As we have shown in Refs.~\cite{Falteralt,Falterneu} the single-hadron attenuation (\ref{eq:multiplicity-ratio}) in the $^{131}$Xe nucleus is larger than for $^{84}$Kr.

To understand the behavior of the double-hadron multiplicity ratio (\ref{eq:double-hadron}) qualitatively we consider a simple absorption model (cf.~\cite{Kopeliovich}). We assume that the leading prehadrons that contribute to the ratio $R_1=N_1|_A/N_1|_D$ start to interact with a cross section $\sigma$ at distance $l_p$ behind the $\gamma^*N$ interaction point $(\vec b,z)$, i.e.
\begin{equation}
	R_1=\frac{1}{A}\intop d^2b\intop_{-\infty}^{\infty}dz\,\rho_A(\vec b,z) \exp\big[-\sigma\intop_{z+l_p}^\infty dz'\rho_A(\vec b,z')\big]\,,
\end{equation}
where $A$ denotes the mass number of the target nucleus and $\rho_A$ its density. If one assumes the production lengths $l_1$ and $l_2$ for the leading and subleading prehadrons, one gets for the double-hadron multiplicity ratio
\begin{equation}
	R_2=\frac{1}{AR_1}\intop d^2b\intop_{-\infty}^{\infty}dz\,\rho_A(\vec b,z) \exp\big[-\sigma\intop_{z+l_1}^\infty dz'\rho_A(\vec b,z')\big]\exp\big[-\sigma\intop_{z+l_2}^\infty dz'\rho_A(\vec b,z')\big]\label{eq:double}\,.
\end{equation}
Even if all production lengths are the same ($l_p=l_1=l_2\equiv l$), $R_2$ will differ from $R_1$:
\begin{equation}
\label{eq:double2}
R_2=\frac{\intop d^2b\intop_{-\infty}^{\infty}dz\,\rho_A(\vec b,z)\exp\big[-2\sigma\intop_{z+l}^\infty dz'\rho_A(\vec b,z')\big]}{\intop d^2b\intop_{-\infty}^{\infty}dz\,\rho_A(\vec b,z)\exp\big[-\sigma\intop_{z+l}^\infty dz'\rho_A(\vec b,z')\big]}\neq R_1\,.
\end{equation}
The numerator and the denominator in Eq.~(\ref{eq:double2}) differ by the absorption cross section in the exponent which for the case of double-hadrons is twice the cross section for the single hadron absorption. Hence, $R_2$ is smaller than one for pure absorption and depends only very weakly on the mass number $A$ of the target nucleus, since the $A$-dependence of the numerator and the denominator are about the same. The dashed line in Fig.~\ref{fig:double} shows the result of our BUU model for a purely absorptive treatment of the FSI. The increase of $R_2(z_2)$ with decreasing $z_2$ is obviously due to the coupled-channel effects: For the interpretation we discard for a moment the constant factors $N_1$ in Eq.~(\ref{eq:double-hadron}) and the factors $N_e$ in Eq.~(\ref{eq:multiplicity-ratio}) which have no influence on the shape of the $z_h$ dependence. The only difference between $N_h(z_h)|_A/N_h(z_h)|_D$ and $N_2(z_2)|_A/N_2(z_2)|_D$ then is that one restricts the detected hadron to the subleading particle in the latter case. If the subleading particle (with energy fraction $z_2$) of the initial $\gamma^*N$ reaction interacts with the nuclear environment, it will produce a bunch of low-energy particles. The {\it new} subleading hadron in the event then has a energy fraction $z_2'<z_2$. As for the usual charged hadron multiplicity spectrum the coupled-channel FSI shuffle strength from the high $z_h$ part to the low $z_h$ part of the spectrum. This is not the case for purely absorptive FSI (dashed line in Fig.~\ref{fig:double}). As one can see, our coupled-channel calculations (solid line) -- using the constituent quark concept (\ref{eq:prehadrons}) and the formation time $\tau_f=0.5$ fm/$c$ -- are again in quantitative agreement with the experimental data apart from the last data point in the $^{84}$Kr data, which indicates a multiplicity ratio $R_2(z_2=0.5)\approx 1$. This behavior cannot be explained within our model.

\section{Summary \& Outlook}
We have presented a model which combines a quantum mechanical treatment of coherence length effects in high-energy (virtual) photon-nucleus interactions with a probabilistic coupled-channel description of hadronic FSI. This allows us to model hadron electro- and photoproduction off complex nuclei at multi GeV energies in a way that goes far beyond simple absorption models. Furthermore, the model can be used with basically the same parameters and physics assumption to describe a broad range of different nuclear reactions such as pion and proton induced reactions as well as heavy-ion-collisions.

In this work we have compared our model results to experimental data on single- and double-hadron multiplicity ratios in deep inelastic positron scattering off nuclear targets at HERMES. We achieve a good description of most experimental data by assuming early interactions of color neutral prehadrons that are produced in the initial $\gamma^*N$ interaction and account for particle absorption {\it and} production in the FSI. However, the model underestimates the increase of high transverse momentum hadrons in $eA$ reactions which may indicate additional partonic effects in the FSI.

We are currently optimizing our model in the HERMES kinematic regime by implementing radiative corrections and \Pythia{} parameters that have been adjusted by the HERMES collaboration \cite{Patty}. Furthermore, we will extend our studies to $eA$ scattering at JLab \cite{Brooks}, EMC \cite{EMC} and EIC \cite{EIC} energies in the future. Since our model works on an event-by-event basis its output can later on be used for various detector studies. In addition we will improve our space-time description of hadronization. In Ref.~\cite{KaiLatest} we have developed a method to extract the complete four-dimensional space-time coordinates of the prehadron production and hadron formation points from the \Jetset{} routines of \Pythia{} on an event-by-event basis. This information will be implemented into our BUU transport model. Furthermore, one can think of considering an additional partonic energy loss prior to color neutralization in our model.

\section{Acknowledgments}
The authors acknowledge valuable discussions with E.~Aschenauer, H.~Blok, M.~D\"uren and P.~Di Nezza. This work was supported by BMBF.



\begin{thebibliography}{99}

\bibitem{Kop} 
B.~Kopeliovich, J.~Nemchik, and E.~Predazzi, 
in {\it Proceedings of the workshop on Future Physics at HERA}, 
edited by G.~Ingelman, A.~De Roeck, R.~Klanner, DESY, 1995/96, vol 2, p. 1038, nucl-th/9607036.

\bibitem{HERMESDIS}
A.~Airapetian et al.~[HERMES Collaboration],
Eur.~Phys.~J.~C {\bf 20}, 479 (2001).
 
\bibitem{HERMESDIS_new}
A.~Airapetian et al.~[HERMES Collaboration],
Phys.~Lett.~B {\bf 577}, 37 (2003).

\bibitem{ArleoWang}
E. Wang and X.-N.~Wang, 
Phys.~Rev.~Lett. {\bf 89}, 162301 (2002);
F. Arleo, 
Eur.~Phys.~J. C {\bf 30}, 213 (2003).

\bibitem{Accardi}
A.~Accardi, V.~Muccifora and H.-J.~Pirner, 
Nucl.~Phys.~A {\bf 720}, 131 (2003).

\bibitem{Falteralt}
T.~Falter, W.~Cassing, K.~Gallmeister and U.~Mosel, 
Phys.~Lett.~B {\bf 594}, 61 (2004).

\bibitem{Falterneu} 
T.~Falter, W.~Cassing, K.~Gallmeister and U.~Mosel,
Phys.~Rev.~C {\bf 70}, 054609 (2004).

\bibitem{Kopeliovich}
B.~Z.~Kopeliovich, J.~Nemchik, E.~Predazzi and A.~Hayashigaki,
Nucl.~Phys.~A {\bf 740}, 211 (2004).

\bibitem{Eff00}
M.~Effenberger and U.~Mosel, 
Phys.~Rev.~C {\bf 62}, 014605 (2000).

\bibitem{Falter}
T.~Falter and U.~Mosel,
Phys.~Rev.~C {\bf 66}, 024608 (2002);
T.~Falter, K.~Gallmeister and U.~Mosel, 
Phys.~Rev.~C {\bf 67}, 054606 (2003), 
Erratum-ibid. {\bf C68}, 019903 (2003).

\bibitem{FalterJLab}
L.~Alvarez-Ruso, {\it et al.}, 
Prog.~Part.~Nucl.~Phys.~({\it in press}), nucl-th/0412084;
T.~Falter, W.~Cassing, K.~Gallmeister, and U.~Mosel,
Acta Phys.~Hung.~A ({\it in press}), nucl-th/0502028.

\bibitem{Gal03}
K.~Gallmeister, C.~Greiner, and Z.~Xu, 
Phys.~Rev.~C {\bf 67}, 044905 (2003).

\bibitem{Cass04}
W.~Cassing, K.~Gallmeister, and C.~Greiner, 
Nucl.~Phys.~A {\bf 735}, 277 (2004);
J.~Phys.~G {\bf 30}, S801 (2004).

\bibitem{PYTHIA}
T.~Sj\"ostrand, {\it et al.},
Comput.~Phys.~Commun.~{\bf 135}, 238 (2001);
T.~Sj\"ostrand, L.~L\"onnblad, S.~Mrenna, PYTHIA 6.2 Physics and Manual,
hep-ph/0108264.

\bibitem{FRITIOF}
Hong Pi,
Comput.~Phys.~Commun.~{\bf 71}, 173 (1992);
B.~Andersson, G.~Gustafson, and Hong Pi,
Z.~Phys.~C {\bf 57}, 485 (1993).

\bibitem{Lund}
B.~Andersson, G.~Gustafson, G.~Ingelman, and T.~Sj\"ostrand,
Phys.~Rep.~{\bf 97}, 31 (1983);
B.~Andersson,
{\it The Lund Model},
Cambridge University Press, 1998.

\bibitem{HSDUrQMD}
W.~Cassing and E.~L.~Bratkovskaya,
Phys.~Rep.~{\bf 308}, 65 (1999);
S.~A.~Bass {\it et al.},
Prog.~Part.~Nucl.~Phys.~{\bf 41}, 225 (1998).

\bibitem{Cas02}
W.~Cassing, E.~L.~Bratkovskaya, and O.~Hansen,
Nucl.~Phys.~A {\bf 707}, 224 (2002).

\bibitem{Eff99c}
M.~Effenberger, E.~L.~Bratkovskaya, and U.~Mosel,
Phys.~Rev.~C {\bf 60}, 44614 (1999).

\bibitem{Nezza}
P.~Di Nezza [HERMES Collaboration],
J.~Phys.~G. {\bf 30}, S783 (2004).

\bibitem{Patty}
Patricia Liebing, Ph.D. Thesis, 2004, University of Hamburg.

\bibitem{Brooks}
W.~K.~Brooks, 
Fizika B {\bf 13}, 321 (2004).

\bibitem{EMC}
J.~Ashman {\it et al.} [EMC],
Z.~Phys.~C {\bf 52}, 1 (1991).

\bibitem{EIC}
A.~Deshpande {\it et al.},
{\it The Electron-Ion Collider: A high luminosity probe of the partonic substructure of nucleons and nuclei},
available electronically from {\it http://www.doe.gov/bridge} (2002).

\bibitem{KaiLatest}
K.~Gallmeister and T.~Falter,
nucl-th/0502015.

\end{thebibliography}
\end{document}